\documentclass[pre,twocolumn]{revtex4}

\usepackage{psfrag}
\usepackage[T1]{fontenc}

\usepackage{times}
\usepackage{ae,aecompl}
\usepackage{graphicx}

\begin{document}

\title{Fractal Structure of Spin Clusters and Domain Walls in
  two-dimensional Ising Model}
\author{Wolfhard Janke and Adriaan M. J.  Schakel}
\affiliation{Institut f\"ur Theoretische Physik, Universit\"at Leipzig,
  Augustusplatz 10/11, 04109 Leipzig, Germany }

\begin{abstract}
  The fractal structure of spin clusters and their boundaries in the
  critical two-dimensional Ising model is investigated numerically.  The
  fractal dimensions of these geometrical objects are estimated by means
  of Monte Carlo simulations on relatively small lattices through
  standard finite-size scaling.  The obtained results are in excellent
  agreement with theoretical predictions and partly provide significant
  improvements in precision over existing numerical estimates. 
\end{abstract}

\date{\today}

\maketitle

\section{Introduction}
The past few years have witnessed a surge in the geometrical approach to
phase transitions.  The prototype of such an approach is percolation
theory \cite{StauferAharony}, which focuses on clusters of (randomly)
occupied sites or bonds on a lattice.  The fractal structure of these
geometrical objects and whether or not a cluster percolates the lattice
are central topics addressed by the theory.  Spin models such as the
$q$-state Potts models can easily be mapped onto percolation theory,
with neighboring spins in the same spin state lumped together in a
cluster.  Generally, the geometrical spin clusters thus constructed do
not percolate at the critical temperature $T_\mathrm{c}$ where the
thermal phase transition takes place.  But by erasing with a certain
temperature-dependent probability bonds between like spins, Fortuin and
Kasteleyn  (FK) \cite{FK} showed that spin clusters can be constructed
for the Potts models that do percolate at $T_\mathrm{c}$ and encode the
thermal critical behavior.  They thus achieved a geometrical description
of the thermal phase transition in these models.  The cluster approach
has been turned into an efficient Monte Carlo algorithm by Swendsen and
Wang \cite{SwendsenWang}, and by Wolff \cite{Wolff}, where not
individual spins are updated, as in local spin updates with, e.g., the
Metropolis algorithm, but entire FK clusters. 

An exception to the rule that geometrical clusters do not percolate at
$T_\mathrm{c}$ is the two-dimensional (2D) $q$-state Potts model.  The
origin of this effect can be understood by extending the pure lattice model to
include vacant sites.  In a Kadanoff block-spin approach, such an
extension is natural as the vacant sites represent disordered blocks
without a majority in any of the spin states, and is essential for
establishing that the phase transition of the pure models changes from
being continuous to first order at $q=4$
\cite{NBRS,CardyNauenbergScalapino}.  In addition to the pure Potts
critical behavior, the site diluted model also displays tricritical
behavior at the same critical temperature $T_\mathrm{c}$.  While the
critical behavior of the pure model is encoded in the FK clusters, the
tricritical behavior is encoded in the geometrical clusters
\cite{StellaVdzandePRL,Vanderzande,geoPotts}.  Both cluster types
percolate at $T_\mathrm{c}$.  With increasing $q$, the critical and
tricritical fixed points, which are characterized by the same central
charge $c$, move together until merging at $q=4$. 

Very recently, cluster boundaries of critical 2D systems have been
studied analytically by means of the so-called stochastic Loewner
evolution, introduced by Schramm \cite{Schramm}.  Various exact
predictions for critical exponents previously conjectured on the basis
of the Coulomb-gas map \cite{denNijs,Nienhuis,SD} and conformal
invariance \cite{Cardyrev} could rigorously be established by this
method (for an overview see Ref.~[\onlinecite{Duplantier02}]). 

In this paper, we numerically investigate the fractal structure of
clusters in the 2D Ising model, corresponding to setting $q=2$.  We
simulate the model on relatively small lattices ($L = 8 \, - \, 512$) with
periodic boundary conditions, and apply standard finite-size scaling to
determine the various fractal dimensions.  In addition to studying the
size or ``mass'' of FK and geometrical clusters, we also examine their
boundaries.  Those of geometrical clusters form the famous Peierls
domain walls \cite{Peierls}, separating spin clusters of opposite
orientation.  In a previous paper \cite{geoPotts}, we simulated these
domain walls directly by considering the high-temperature representation
of the model.  By duality, the high-temperature graphs, which are
closed, are domain walls on the dual lattice.  The closed graph
configurations were generated by means of a Metropolis update algorithm,
involving single plaquettes. 

Other recent numerical studies of the geometrical structure of 2D Potts
models were reported in Refs.~[\onlinecite{Fortunato,AABRH}].  Our
results for the fractal dimensions are in excellent agreement with
theoretical predictions
\cite{Stanley,SD,StellaVdzandePRL,VdzandeStellaJP,Duplantier00}, and
provide in particular for the FK and geometrical clusters a considerable
improvement in precision over the estimates obtained in
Refs.~[\onlinecite{Fortunato,AABRH}]. 

The rest of the paper is structured as follows.  The next section
summarizes the necessary theoretical background.  Numerical results are
presented in Sec.~\ref{sec:MC}, followed by concluding remarks in
Sec.~\ref{sec:conclude}.

\section{Fractal Structures}
The fractal properties of spin clusters and boundaries, which are
clusters themselves, are described by a straightforward extension of
ordinary percolation theory \cite{StauferAharony}.  Asymptotically,
cluster distributions $\ell_n$ take a general form
\begin{equation} 
\label{ell}
\ell_n \sim n^{- \tau} \exp(- \theta n),
\end{equation} 
consisting of two factors: (i) an entropy factor, which measures the
number of ways a cluster of size $n$ can be embedded in the lattice, and
(ii) a Boltzmann weight, which suppresses large clusters when $\theta$
is finite.  Clusters proliferate and percolate the lattice when $\theta$
tends to zero.  The vanishing is characterized by an exponent $\sigma$
as $\theta \propto |T-T_{\rm p}|^{1/\sigma}$, where $T_{\rm p}$ denotes
the percolation temperature.  As explained in the Introduction, the
percolation thresholds of both FK and geometrical clusters coincide with
the thermal critical temperature of the 2D Ising model.  The entropy
exponent $\tau$ determines the fractal structure of the geometrical
objects.  Rather than extracting this exponent directly from the
asymptotic behavior of a distribution at the percolation threshold,
where the distribution becomes algebraic, it is expedient to extract it
from derived quantities such as the percolation strength $P_\infty$,
giving the fraction of sites in the largest cluster, and the average
cluster size \cite{StauferAharony}
\begin{equation} 
\label{chi}
\chi = \frac{\sum_n n^2 \ell_n}{\sum_n n \ell_n}. 
\end{equation} 
Since every site belongs to some geometrical and some FK cluster, the
denominator in Eq.~(\ref{chi}) adds up to the total number of sites for
these clusters.  Close to the percolation threshold, the observables
obey the finite-size scaling laws \cite{BinderHeermann}
\begin{equation} 
\label{finitess}
P_\infty = L^{-\beta/\nu} \, {\sf P}(L/\xi), \quad
\chi = L^{\gamma/\nu} \, {\sf X} (L/\xi),
\end{equation} 
where $L$ is the lattice size and $\xi$ the correlation length whose
divergence at criticality is governed by the exponent $\nu$.  The ratios
$\beta/\nu$ and $\gamma/\nu$ are given in terms of $\tau$ as
\cite{StauferAharony}
\begin{equation} 
\frac{\beta}{\nu} = d \frac{\tau-2}{\tau-1},  \quad \frac{\gamma}{\nu} =  
d \frac{3-\tau}{\tau-1}, 
\end{equation} 
with $d$ the dimensionality of the lattice.  The fractal dimension $D$,
which is also determined solely by the entropy exponent $\tau$, is related to
these exponents via \cite{StauferAharony}
\begin{equation}  
\label{Dbg}
D = \frac{d}{\tau-1} = d - \frac{\beta}{\nu} = \frac{1}{2} \left(d +
\frac{\gamma}{\nu} \right). 
\end{equation} 

Generically, two (and only two) different cluster boundaries can be
identified \cite{GrossmanAharony}: the hull (H) and the external
perimeter (EP), where the second can be understood as a smoother version
of the first.  For 2D FK clusters, the two boundaries are in one-to-one
correspondence, with their fractal dimensions satisfying the relation
\cite{Duplantier00}
\begin{equation} 
\label{Drel}
(D^\mathrm{FK}_\mathrm{H}-1) (D^\mathrm{FK}_\mathrm{EP}-1) =
\textstyle\frac{1}{4}. 
\end{equation}
The map transforming one FK boundary dimension into the other conserves
the central charge $c$, which may be parametrized as
\cite{Nienhuis,Cardyrev}
\begin{equation} 
\label{conf}
c = 1 - \frac{6(1-\bar\kappa)^2}{\bar\kappa} = 13 - 6 
\left(\bar\kappa +\frac{1}{\bar\kappa}\right),
\end{equation}   
where $2 \ge \bar\kappa \ge 1$ parametrizes the two-dimensional
$q$-state Potts models
\begin{equation}
\label{Potts_branch} 
\sqrt{q} = - 2 \, \cos(\pi/\bar\kappa),
\end{equation} 
with $0\leq q \leq 4$.  In terms of $\bar\kappa$, the fractal dimensions
of the FK boundaries \cite{SD,Duplantier00} can be expressed as
\begin{equation} 
\label{DPotts}
D^\mathrm{FK}_\mathrm{H} = 1 + \frac{\bar\kappa}{2}, \quad
D^\mathrm{FK}_\mathrm{EP} = 1 + \frac{1}{2\bar\kappa}  ,
\end{equation} 
while the central charge conserving map corresponds to letting $\bar\kappa
\to 1/\bar\kappa$.  These explicit forms are seen to satisfy the duality
relation (\ref{Drel}).  With the scaling relations (\ref{Dbg}), the
critical exponent ratios characterizing the FK boundaries become
\begin{equation} 
\gamma^\mathrm{FK}_\mathrm{H}/\nu = \bar\kappa, \quad
\gamma^\mathrm{FK}_\mathrm{EP}/\nu = 1/\bar\kappa,
\end{equation} 
where a single correlation length with exponent $\nu$ is assumed.  It
thus follows that the two FK boundary sizes scale with inverse
exponents:
\begin{equation} 
\chi^\mathrm{FK}_\mathrm{H} \sim L^{\bar \kappa}, \quad
\chi^\mathrm{FK}_\mathrm{EP} \sim L^{1/\bar\kappa}. 
\end{equation} 

In contrast to FK clusters, geometrical clusters are characterized by
only one boundary dimension, i.e., the fractal dimensions
$D^\mathrm{G}_\mathrm{H}$ and $D^\mathrm{G}_\mathrm{EP}$ of the hull and
external perimeter coincide, $D^\mathrm{G}_\mathrm{H} =
D^\mathrm{G}_\mathrm{EP}$.  Such cases are signaled by a negative
fractal dimension of the red sites \cite{Stanley}, sites that, when
removed, lead to a splitting of the cluster into disconnected parts. 

The central charge conserving map $\bar\kappa \to 1/\bar\kappa$
transforming the hull dimension $D^\mathrm{FK}_\mathrm{H}$ of FK
clusters into that of their external perimeters,
$D^\mathrm{FK}_\mathrm{EP}$, also maps it onto the hull dimension of
geometrical clusters, implying \cite{Duplantier00}
\begin{equation} 
\label{DH-DEP}
D^\mathrm{FK}_\mathrm{EP} = D^\mathrm{G}_\mathrm{H} . 
\end{equation} 
This relation is remarkable as it involves the two different boundary
types.  In the context of uncorrelated percolation
\cite{GrossmanAharony}, the hull of a cluster in 2D is traced out by a
directed random walker constrained to move on the cluster only, whereas
the external perimeter is traced out by a walker constrained to move
around the hull on sites neighboring, but not belonging to, that cluster. 

\begin{figure}
\begin{center}
\includegraphics[width=8.0cm]{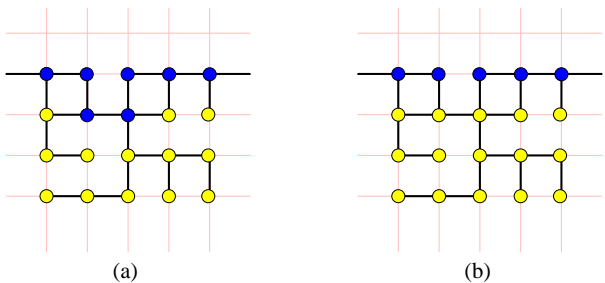}
\end{center}
\caption{(Color online) Part of a single FK cluster of nearest neighbor sites (filled circles) 
  connected by bonds (black links). (a) Sites belonging to the
  hull (dark filled circles) are found by allowing the random walker
  tracing out the boundary to move only over set bonds. 
  (b) Sites belonging to the external perimeter (dark filled
  circles) are found by allowing the random walker to move to a
  nearest neighbor on the cluster irrespective of whether the connecting
  bond is set or not.  The external perimeter, which contains two sites
  less than the hull for this boundary segment, is therefore a smoother
  version of the hull. 
  \label{fig:rw_algo}}
\end{figure}

To numerically verify relation (\ref{DH-DEP}), we wish to treat external
perimeters of FK clusters and hulls of geometrical clusters in the same
manner.  To this end we apply the same algorithm used to trace out
geometrical hulls to find the FK external perimeters, where it is
recalled that FK clusters differ from geometrical clusters in that with
a prescribed temperature-dependent probability bonds are erased.  The
difference between tracing out FK hulls and external perimeters then
reduces to (see Fig.~\ref{fig:rw_algo}) allowing the random walker to
move to a nearest neighbor site on the FK boundary only provided the
connecting bond is set (hull) or always (external perimeter). 

To conclude this section we list in Table \ref{table:exact} the
predicted exact values
\cite{Stanley,SD,StellaVdzandePRL,VdzandeStellaJP,Duplantier00} for the
various fractal dimensions and corresponding critical exponents we wish
to determine numerically. 
\begin{table}
\begin{tabular}{c|ccccccc}
\hline \hline & & & & & & & \\[-.2cm] 
 & $D_\mathrm{C}$ & $\gamma_\mathrm{C}/\nu$ &
$\beta_\mathrm{C}/\nu$ & $D_\mathrm{H}$ & $\gamma_\mathrm{H}/\nu$ &
$D_\mathrm{EP}$ & $\gamma_\mathrm{EP}/\nu$ \\[.3cm]
\hline & & & & & & & \\[-.3cm]
FK & $15/8$ & $7/4$ & $1/8$ & $5/3$ &
$4/3$ & $11/8$ & $3/4$   \\[.1cm]
G & $187/96$ & $91/48$ & $5/96$ & $11/8$ &
$3/4$ &  & \\[.1cm] \hline \hline
\end{tabular}
\caption{Predicted values for the fractal dimensions with the
corresponding critical exponents characterizing Fortuin-Kasteleyn (FK)
and geometrical (G) clusters (C), their hulls (H), and external
perimeters (EP). 
\label{table:exact}}
\end{table}

\section{Monte Carlo Simulations}
\label{sec:MC}
The simulation data was collected on square lattices of linear size
$L=8, 10, 12, 14$, $16, 20, 24, 32$, $40, 48, 64, 90$, $128, 180, 256,
360$, and $512$ with periodic boundary conditions, using the
Swendsen-Wang cluster algorithm \cite{SwendsenWang} in about $5 \times
10^4$ measurements at the critical temperature $T_\mathrm{c} =
2/\ln(1+\sqrt{2})$, every $\tau$th sweep of the lattice, where $\tau$
denotes the autocorrelation time (rounded off to the next largest
integer).  The value of $\tau$ for the various lattice sizes $L$ was
estimated from the energy time series to vary from $\tau\approx4$ for
$L=8$ to $\tau\approx9$ for $L=512$.  We have chosen the energy time
series here as it generally leads to a conservative estimate of the
autocorrelation time for cluster algorithms.  A total of $5 \times 10^3$
lattice sweeps were used for equilibration.  Statistical errors were
estimated by means of jackknife binning \cite{Efron}. 

\subsection{Fractal Dimensions}
Tables \ref{table:SW_cluster}, \ref{table:geo_cluster},
\ref{table:SW_hull}, and \ref{table:SW_ep} summarize the values obtained
for the critical exponents of the FK and geometrical clusters, as well
as of the FK hulls and external perimeters.  Where $\gamma/\nu$ and
$\beta/\nu$ are listed, both are measured independently by considering
the average cluster size $\chi$, which gives $\gamma/\nu$ according to
Eq.~(\ref{finitess}) with ${\sf X} (0)= \mbox{const.}$ and the
percolation strength $P_\infty$, which gives $\beta/\nu$ according to
Eq.~(\ref{finitess}) with ${\sf P}(0)=\mbox{const}$.  The data were
fitted using the least-squares Marquardt-Levenberg algorithm. 

While including the percolating clusters when considering the mass of
the cluster, we ignore them in tracing out cluster \textit{boundaries}. 
Because of the finite lattice size, large percolating clusters have
abnormal small (external) boundaries, so that including them would lead
to a distortion of the hull distribution.  Moreover, the
Grossman-Aharony algorithm \cite{GrossmanAharony} we use to trace out
the cluster boundaries generally fails on a percolating cluster as its
boundary not necessarily forms a single closed loop any longer.  With
the percolating clusters ignored, the boundary exponents
$\beta_\mathrm{H,EP}/\nu$ cannot be determined, while the summation in
the expression (\ref{chi}) for the average cluster boundary size is
restricted to nonpercolating clusters.

\subsubsection{FK Clusters}
In Ref.~[\onlinecite{AABRH}], the value $D^\mathrm{FK}_\mathrm{C} =
1.87(1)$ compared to the exact result \cite{Stanley}
$D^\mathrm{FK}_\mathrm{C}=15/8=1.875$ was reported for the fractal
dimension of FK clusters.  It was obtained on a single, but very large
lattice ($L=2^{12}=4096$) with both open and periodic boundary
conditions by considering the number of bond clusters as a function of the
radius of gyration.  The authors observed a slow and complex approach to
the asymptotic behavior and therefore included corrections to scaling to
arrive at their numerical estimates.  To limit the number of fit
parameters, they scanned the fractal dimensions in the neighborhood of
the predicted values, and completely fixed the values of the correction
exponents to the theoretically predicted ones. This left them with
still four parameters to fit.  Error bars on the values of the fractal
dimensions were determined as the range where $\chi^2/\mathrm{d.o.f.} 
<2$. 

In this study, where we use different lattice sizes and consider not
the bonds, but the sites in a cluster, we find a simple approach to the
asymptotic behavior with very small corrections to scaling in the
observables required to determine $D^\mathrm{FK}_\mathrm{C}$ all the way
down to the smallest lattice considered ($L=8$).  We can therefore apply
finite-size scaling without correction terms.  Since the fits involve
only two parameters, no exponents need to be fixed beforehand.  To
minimize the effect of unavoidable (small) corrections to scaling, we
pick the fit over the largest lattice sizes given in
Table~\ref{table:SW_cluster}, i.e., over the range $64 - 512$, leading
to
\begin{equation} 
D^\mathrm{FK}_\mathrm{C}=1.8753(6)
\end{equation} 
with $\chi^2/\mathrm{d.o.f} = 0.61$ from the average cluster-size data and
\begin{equation} 
D^\mathrm{FK}_\mathrm{C}=1.8752(8)
\end{equation} 
with $\chi^2/\mathrm{d.o.f} = 0.47$ from the percolation-strength data. 
Both estimates are well within one standard deviation from the exact
prediction \cite{Stanley} $D^\mathrm{FK}_\mathrm{C}=15/8=1.875$. 

A possible explanation for the improved accuracy we achieved over
Ref.~[\onlinecite{AABRH}], although working on smaller lattices, may be
that in that study clusters touching the boundary were ignored in the
measurements.  This set includes all the percolating clusters.  As will
be illustrated in the next subsection, omitting percolating clusters
can lead to strong corrections to scaling. 

\subsubsection{Geometrical Clusters}
In Ref.~[\onlinecite{Fortunato}], the values
$\gamma^\mathrm{G}_\mathrm{C}/\nu = 1.901(11)$ and
$\beta^\mathrm{G}_\mathrm{C}/\nu=0.052(2)$ were reported for geometrical
clusters.  These results were obtained on lattices ranging in size from
$L=600$ to $2000$, i.e., again much larger than the ones considered by
us.  Instead of using periodic boundary conditions, as we did, free
boundary conditions were adopted.  Another difference from our approach
is that percolating clusters were excluded in
Ref.~[\onlinecite{Fortunato}]. 

Our estimates, obtained from the largest lattice sizes listed in
Table~\ref{table:geo_cluster}, i.e., from the interval 
$64 - 512$, are
\begin{equation} 
D_\mathrm{C}^\mathrm{G} = 1.9476(3)
\end{equation}
with $\chi^2/\mathrm{d.o.f} = 0.44$ from the average cluster-size data and
\begin{equation} 
D_\mathrm{C}^\mathrm{G} = 1.9473(4)
\end{equation}
with $\chi^2/\mathrm{d.o.f} = 0.23$ from the percolation-strength data. 
Both are in excellent agreement with the exact prediction
\cite{StellaVdzandePRL} $D^\mathrm{G}_\mathrm{C}=187/96=1.9479 \dots$. 
\begin{table}
\begin{tabular}{lllll}
\hline \hline &  &  \\[-.2cm] 
Fit Interval & \hspace{10.pt} $\gamma^\mathrm{FK}_\mathrm{C}/\nu$ &
$\chi^2/$d.o.f. &
\hspace{10.pt} $\beta^\mathrm{FK}_\mathrm{C}/\nu$ & $\chi^2/$d.o.f.  \\[.3cm]
\hline & & & &\\[-.3cm]

\hspace{1.ex}$8 - 256$ & 1.7512(6)  & 0.89 & 0.1244(4)  & 0.74 \\
            $16 - 256$ & 1.7507(8)  & 1.02 & 0.1246(5)  & 0.95 \\ 
            $32 - 256$ & 1.7500(12) & 1.39 & 0.1249(8)  & 1.31 \\    
            $40 - 256$ & 1.7507(13) & 1.43 & 0.1246(9) & 1.43 \\    
            $64 - 256$ & 1.7496(21) & 0.52 & 0.1254(14)  & 0.33 \\

& &  & & \\

\hspace{1.ex}$8 - 512$ & 1.7511(4)  & 0.86 & 0.1244(3)  & 0.71 \\
             $16 - 512$ & 1.7507(6)  & 0.95 & 0.1246(4)  & 0.88 \\
             $32 - 512$ & 1.7505(8)  & 1.23 & 0.1247(5)  & 1.14 \\    
             $40 - 512$ & 1.7509(9)  & 1.20 & 0.1245(6)  & 1.17 \\
             $64 - 512$ & 1.7505(12) & 0.61 & 0.1248(8)  & 0.47 \\[.1cm] \hline
     exact  & 1.7500 = 7/4 & &  0.1250 = 1/8 & \\[.1cm] \hline \hline
\end{tabular}
\caption{FK clusters. 
\label{table:SW_cluster}}
\end{table}

\begin{figure}
\begin{center}
\includegraphics[width=8.0cm]{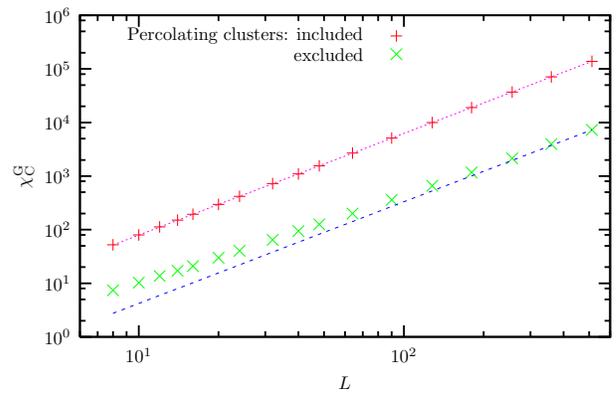}
\end{center}
\caption{(Color online) Log-log plot of the average size
  $\chi^\mathrm{G}_\mathrm{C}$ of critical geometrical clusters as a
  function of the lattice size $L$.  Statistical error bars are smaller
  than the symbol size in the figure.  The straight lines proportional to 
  $L^{91/48}$ are put through the data points at $L=512$ by hand to
  demonstrate the corrections to scaling for smaller lattice sizes when
  percolating clusters are excluded, and the absence thereof when they
  are included in the measurements. 
  \label{fig:compare}}
\end{figure}

In Fig.~\ref{fig:compare} we show our data for the average cluster size
obtained by including all clusters and compare it with the data obtained
with percolating clusters excluded from the measurements, similarly to
what was done in Ref.~[\onlinecite{Fortunato}].  While virtually absent
in the former, corrections to scaling are present in the latter case. 
This may explain why, although working on smaller lattices, we obtained
much better estimates than in Ref.~[\onlinecite{Fortunato}].  We found
similar corrections to scaling when instead of percolating clusters, the
largest cluster in each measurement was excluded, as is common in random
percolation \cite{BinderHeermann}.

\begin{table}
\begin{tabular}{lllll}
\hline \hline & & & & \\[-.2cm] 
Fit Interval & \hspace{10.pt} $\gamma^\mathrm{G}_\mathrm{C}/\nu$ &
$\chi^2/$d.o.f.& \hspace{10.pt} $\beta^\mathrm{G}_\mathrm{C}/\nu$ &
$\chi^2/$d.o.f.  \\[.3cm]
\hline & & & & \\[-.3cm]

\hspace{1.ex}$8 - 256$ & 1.8941(3) &  1.16 & 0.0532(2) & 0.83 \\
            $16 - 256$ & 1.8944(4) &  1.39 & 0.0530(3) & 0.94 \\ 
            $32 - 256$ & 1.8946(5) &  1.90 & 0.0528(4) & 1.24 \\    
            $40 - 256$ & 1.8950(6) &  1.78 & 0.0526(4) & 1.09 \\    
            $64 - 256$ & 1.8949(9) &  0.47 & 0.0527(7) & 0.18 \\

& &  & & \\

\hspace{1.ex}$8 - 512$ & 1.8943(2) &  1.29 & 0.0531(2) & 0.89 \\
            $16 - 512$ & 1.8946(3) &  1.34 & 0.0529(2) & 0.88 \\
            $32 - 512$ & 1.8949(4) &  1.60 & 0.0528(3) & 1.02 \\    
            $40 - 512$ & 1.8951(4) &  1.39 & 0.0526(3) & 0.86 \\
            $64 - 512$ & 1.8951(5) &  0.44 & 0.0527(4) & 0.23 \\ [.1cm] \hline
      exact & 1.8958 = 91/48 & &  0.0521 = 5/96 & \\[.1cm] \hline \hline
\end{tabular}
\caption{Geometrical clusters. 
\label{table:geo_cluster}}
\end{table}

\subsubsection{FK Hulls}
Surprisingly, the results for the hulls of FK clusters given in Table
\ref{table:SW_hull} show a clear tendency to the predicted value
\cite{SD} $\gamma^\mathrm{FK}_\mathrm{H}/\nu = 4/3$, corresponding to
$D^\mathrm{FK}_\mathrm{H}=5/3 = 1.6666 \dots$, when restricting the
fitting window to increasingly \textit{smaller} lattice sizes.  For
example, for the interval $8-48$ we find
\begin{equation} 
D^\mathrm{FK}_\mathrm{H}= 1.665(3)
\end{equation} 
with $\chi^2/\mathrm{d.o.f} = 0.79$, indicating a good fit.  This
estimate, which should be compared with the estimate $1.66(1)$ given in
Ref.~[\onlinecite{AABRH}], is within one standard deviation from the
exact prediction. 
\begin{figure}
\begin{center}
  \includegraphics[width=8.0cm]{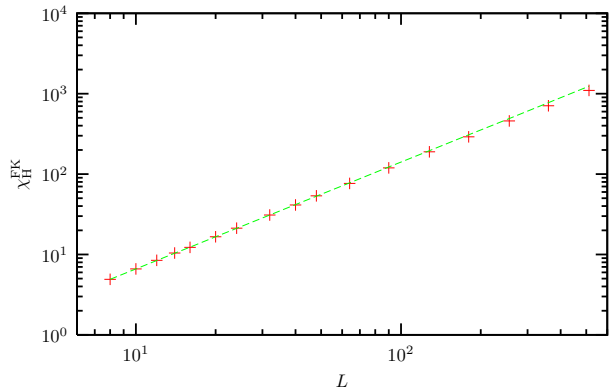}
\end{center}
\caption{(Color online) Log-log plot of the average hull size
  $\chi^\mathrm{FK}_\mathrm{H}$ of critical FK clusters as a
  function of the lattice size $L$.  Statistical error bars are smaller
  than the symbol size in the figure.  The straight line
  $0.310 \, L^{1.329}$ is obtained through a two-parameter fit in the
  interval $8 - 48$. 
\label{fig:SW_hull}}
\end{figure}
From Fig.~\ref{fig:SW_hull} we see that the average FK hull size
measured on larger lattices falls below the expected value extrapolated
from smaller lattices.  We have not been able to determine the cause of
this behavior.  In fact, when fitting not at the low end but at the
high end of the lattice sizes considered, we obtain fits of comparable
quality, but the estimate for the exponent converges to a value
significantly below the predicted one (see top part of Table
\ref{table:SW_hull}). 

\begin{table}
\begin{tabular}{lllll}
\hline \hline & & \\[-.2cm] 
Fit Interval & \hspace{10.pt} $\gamma^\mathrm{FK}_\mathrm{H}/\nu$ &
$\chi^2/$d.o.f. \\[.3cm]
\hline & & \\[-.3cm]
\hspace{1.1ex}$8 - 512$  & 1.304(2) &  4.71 \\
             $16 - 512$ & 1.298(2)  & 3.65 \\
             $32 - 512$ & 1.288(3)  & 1.57 \\
             $40 - 512$ & 1.285(4)  & 1.36 \\    
             $64 - 512$ & 1.281(5)  & 0.62 \\

& &  \\

    \hspace{1.ex}$8 - 256$  & 1.311(2) &  2.88 \\
    \hspace{1.ex}$8 - 128$  & 1.318(3) &  1.50 \\    
    \hspace{1.ex}$8 - \hspace{1.ex}90$   & 1.321(3) &  1.29 \\
    \hspace{1.ex}$8 - \hspace{1.ex}64$   & 1.325(4) &  1.10 \\
    \hspace{1.ex}$8 - \hspace{1.ex}48$   & 1.329(5) &  0.79 \\[.1cm] \hline
exact  & 1.333 = 4/3  & \\[.1cm] \hline \hline
\end{tabular}
\caption{Hulls of FK clusters. 
\label{table:SW_hull}}
\end{table}

\begin{table}
\begin{tabular}{lllll}
\hline \hline & & \\[-.2cm] 
Fit Interval & \hspace{10.pt} $\gamma^\mathrm{FK}_\mathrm{EP}/\nu$ &
$\chi^2/$d.o.f. \\[.3cm]
\hline & & \\[-.3cm]

    $24 - 256$ & 0.763(3)  & 4.32 \\
    $32 - 256$ & 0.758(3)  & 3.10 \\
    $40 - 256$ & 0.755(4)  & 3.24 \\    
    $64 - 256$ & 0.748(6)  & 4.13 \\

& &  \\

    $24 - 512$ & 0.752(2) & 7.97 \\
    $32 - 512$ & 0.747(2) & 5.73 \\
    $40 - 512$ & 0.744(3) & 5.29 \\    
    $64 - 512$ & 0.736(4) & 4.37 \\[.1cm] \hline

exact  & 0.750 = 3/4  & \\[.1cm] \hline \hline
\end{tabular}
\caption{External perimeters of FK clusters. 
\label{table:SW_ep}}
\end{table}

\subsubsection{Geometrical Hulls}
\begin{figure}
\begin{center}
\includegraphics[width=8.0cm]{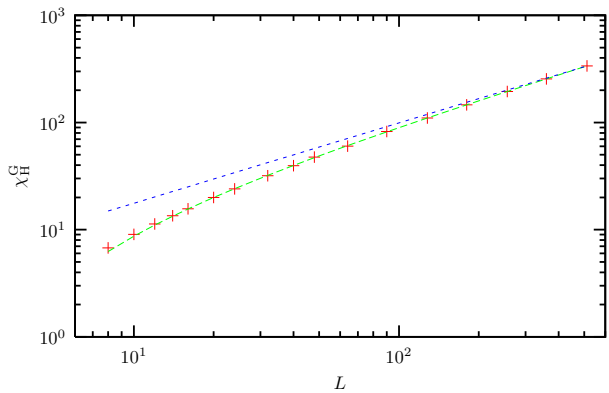}
\end{center}
\caption{(Color online) Log-log plot of the average hull size
  $\chi^\mathrm{G}_\mathrm{H}$ of critical geometrical clusters as a
  function of the lattice size $L$.  Statistical error bars are smaller
  than the symbol size in the figure.  The straight line proportional 
   $L^{3/4}$ is put through the data point at $L=512$ by hand to
  demonstrate the strong corrections to scaling for smaller lattice
  sizes.  A three-parameter fit in the interval $16 - 512$ gives the
  value $w = 0.54(3)$ for the effective correction-to-scaling exponent. 
  \label{fig:geo_hull}}
\end{figure}

As for the clusters' mass when disregarding percolating clusters, we
observe strong corrections to scaling for the hulls of geometrical
clusters (see Fig.~\ref{fig:geo_hull}). This is different from what we
found using the plaquette update to directly simulate the hulls of the
spins on the dual lattice \cite{geoPotts}, where these corrections were
virtually absent (see Fig.~11 of that paper), allowing us to obtain
precise estimates for the critical exponents on relatively small
lattices.  In that study, the largest hull was omitted in each
measurement.

\begin{figure}
\begin{center}
\includegraphics[width=8.0cm]{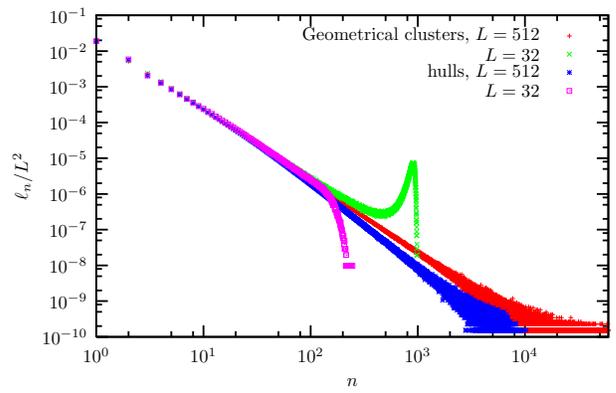}
\end{center}
\caption{(Color online) Log-log plot of the (normalized) geometrical cluster and hull
  distributions at $T_\mathrm{c}$ on the largest lattice considered 
  ($L=512$)  and on 
  a relatively small lattice
  ($L=32$).    The number of
  measurements taken on the largest lattice was about $5 \times 10^4$ as in most part of this
  paper.  Statistical error bars are omitted from the data points for
  clarity.  On the smaller lattice, about $5 \times 10^5$
  measurements, which is an order of magnitude more than used
  in the rest of the paper,  were taken to achieve good statistics. Here, the
  statistical error bars are smaller than the symbol size in
  the figure.  
  \label{fig:geo_dis}}
\end{figure}

To understand the strong corrections to scaling found here for the
geometrical hulls, we depict in Fig.~\ref{fig:geo_dis} the geometrical
cluster and the corresponding hull distributions for $L=32$ normalized
to the volume $L^2$.  The bump at the tail of the cluster distribution
is due to the finite size of the lattice, with percolating clusters
gulping up smaller ones reached by crossing lattice boundaries.  The
subsequent sharp drop-off arises because of the limited number of
lattice sites available. 

Initially, as Fig.~\ref{fig:geo_dis} clearly shows, the hull
distribution follows more or less the cluster distribution.  This is a
common feature of all boundary distributions considered.  The relatively
early drop-off of the hull distribution is because we omit percolating
clusters when tracing out cluster boundaries.  As a result, the average
hull size is underestimated and the data points in
Fig.~\ref{fig:geo_hull} are below the expected line extrapolated from
larger lattice sizes.  With increasing lattice size, the effect becomes
smaller (see Fig.~\ref{fig:geo_dis}, where also the distributions for
$L=512$ are included) and the data points approach the expected
asymptotic scaling, corresponding to \cite{VdzandeStellaJP}
$\gamma^\mathrm{G}_\mathrm{H}/\nu=3/4$, and
$D^\mathrm{G}_\mathrm{H}=11/8$.  Figure~\ref{fig:geo_dis} shows in
addition that the asymptotic behavior of the hull distribution sets in
for relatively large hull sizes ($n \gtrsim 100$).  On smaller lattices,
the asymptotic behavior can therefore simply not be probed, explaining
the strong corrections in Fig.~\ref{fig:geo_hull}. 

To see if our data are at least consistent with the theoretical
prediction, we account for corrections to scaling by fitting the average
hull-size data to the form
\begin{equation} 
\chi^\mathrm{G}_\mathrm{H} = a L^{\gamma^\mathrm{G}_\mathrm{H}/\nu}\left(1-
b L^{-w}\right),
\end{equation} 
with an effective correction-to-scaling exponent $w$.  We fix
$\gamma^\mathrm{G}_\mathrm{H}/\nu=3/4$ to the predicted value, leaving
us with three parameters ($w$ and the two amplitudes $a$ and $b$)
to fit.  For the interval $16 - 512$ we obtain
\begin{equation} 
w = 0.54(3)
\end{equation} 
with $\chi^2/\mathrm{d.o.f} = 1.60$, indicating a reasonable fit (see
Fig.~\ref{fig:geo_hull}) and therefore consistency with the theoretical
prediction for $D^\mathrm{G}_\mathrm{H}$. 

\subsubsection{FK External Perimeters}
\begin{figure}
\begin{center}
\includegraphics[width=8.0cm]{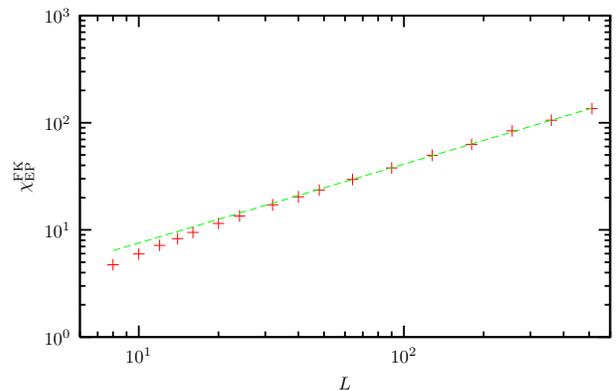}
\end{center}
\caption{(Color online) Log-log plot of the average external perimeter size
  $\chi^\mathrm{FK}_\mathrm{EP}$ of FK
  clusters at $T_\mathrm{c}$  as a function of the lattice size $L$.  Statistical error bars
  are smaller than the symbol size in the figure.  The straight line
  $1.388 \, L^{0.736}$ is obtained through a two-parameter fit in the
  interval $64 - 512$.  Note the corrections to scaling for smaller
  lattice sizes. 
  \label{fig:SW_ep}}
\end{figure}

The corrections to scaling are less pronounced for the external
perimeters of FK clusters as they are generally smaller than geometrical
clusters and thus less likely to percolate (see Fig.~\ref{fig:SW_ep}). 
Also the asymptotic behavior is reached earlier than for geometrical
hulls.  The smallness of the corrections allows us to obtain reasonable
fits for the average external perimeter size.  Our result obtained from
the fitting interval $64 - 512$ in Table~\ref{table:SW_ep} yields
the fractal dimension
\begin{equation} 
D^\mathrm{FK}_\mathrm{EP} = 1.368(2)
\end{equation}  
with $\chi^2/\mathrm{d.o.f} = 4.37$.  This estimate is compatible with
the exact prediction \cite{Duplantier00} $D^\mathrm{FK}_\mathrm{EP} =
11/8 = 1.375$, and improves the estimate $D^\mathrm{FK}_\mathrm{EP} =
1.36(2)$ reported in Ref.~[\onlinecite{AABRH}] by about one order of
magnitude. 

\subsection{Distributions}
\begin{figure}
\begin{center}
\includegraphics[width=8.0cm]{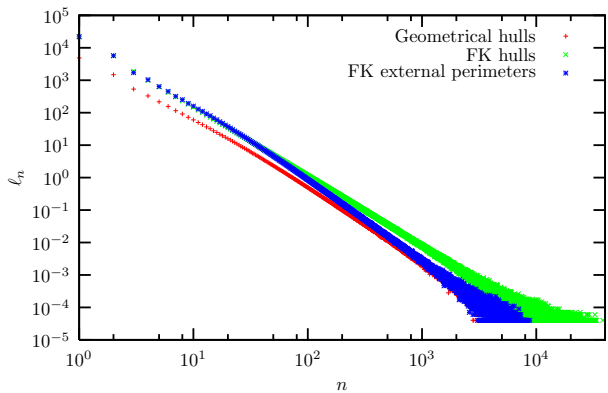}
\end{center}
\caption{(Color online) Log-log plot of the distributions of the three different
  boundaries at $T_\mathrm{c}$ for $L=512$. Statistical error bars are
  omitted from the data points for clarity.  The FK external perimeter
  distribution initially follows the FK hull distribution before at
  around $n \approx 100$ it crosses over to its asymptotic behavior
  which it shares with the geometrical hulls. 
\label{fig:boundaries}}
\end{figure}

In Fig.~\ref{fig:boundaries}, the distributions of the three different
boundaries studied are plotted for $L=512$ to show the crossover of the
external perimeters of FK clusters.  Starting similarly to the FK hull
distribution, the FK external perimeter distribution asymptotically
approaches that of the geometrical hulls, in accordance with relation
(\ref{DH-DEP}).  In other words, the FK external perimeter distribution
interpolates between that of the FK (for small $n$) and geometrical
hulls (asymptotically for large $n$). 

\begin{figure*}
\includegraphics[width=8.0cm]{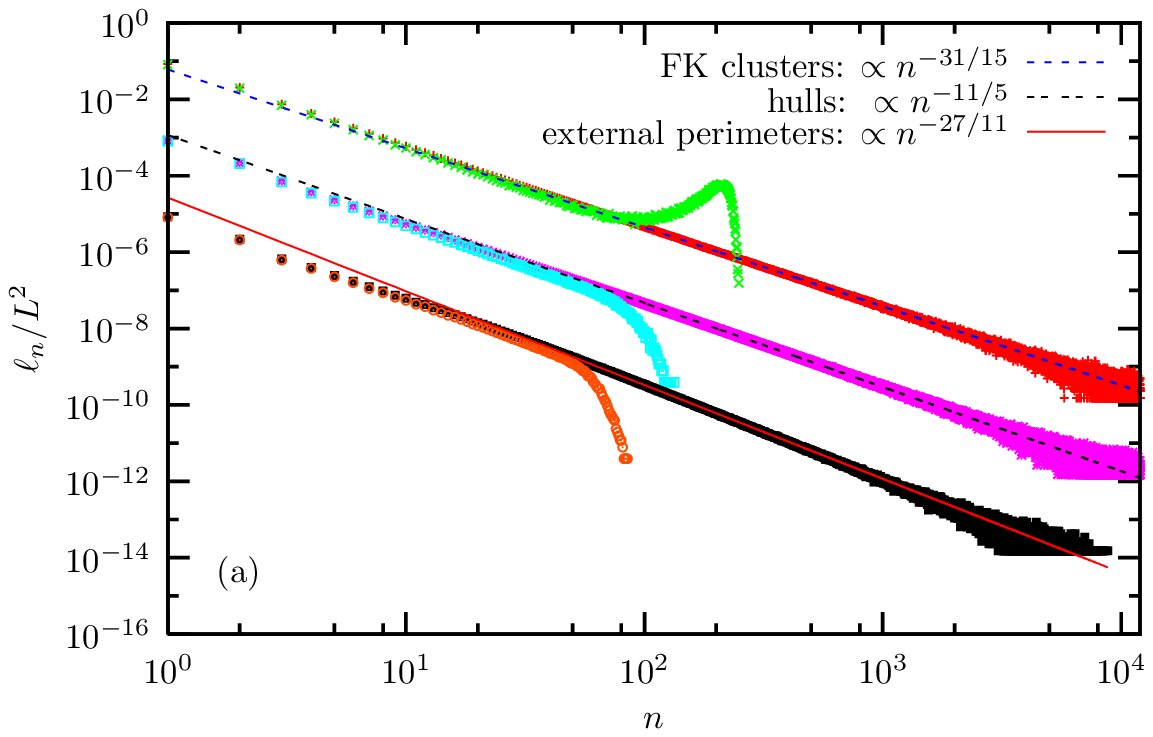} \hspace{1.cm}
\includegraphics[width=8.0cm]{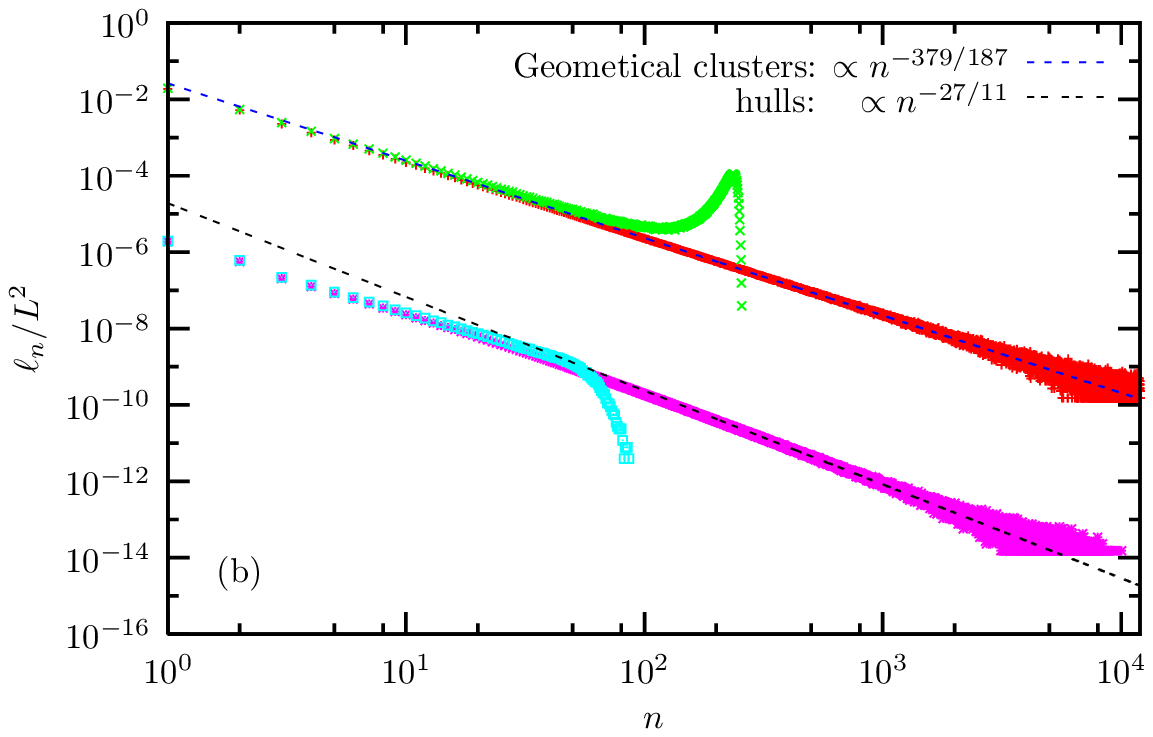} 
\caption{(Color online) Log-log plot of the critical distributions studied for $L=16$ 
(short curves) and $512$ (long curves).  (a) The (normalized) 
distributions of FK clusters and of their hulls 
and external perimeters.  For clarity, the latter two are shifted downward 
by two decades each. 
(b) The (normalized) distributions of geometrical clusters 
and of their hulls.  
The latter is shifted downward by four decades.  Statistical error bars are 
also for clarity omitted from the data
  points. The straight lines are obtained through one-parameter fits
  with the slopes fixed to the predicted values.  To achieve good
  statistics for the $L=16$ lattice, about $5 \times 10^5$ measurements
  were taken--an order of magnitude more than used for the $L=512$
  lattice. 
  \label{fig:distros}}
\end{figure*}

Figure \ref{fig:distros} summarizes all the cluster and boundary
distributions studied for $L=16$ and $512$.  The distributions are
normalized to the volume $L^2$.  Upon increasing the lattice size, the
normalized distributions tend to a universal curve.  The slow approach
to the asymptotic form of the geometrical hull distribution (and to a
lesser extent that of the FK external perimeter distribution), with the
associated strong corrections to scaling we observed for these objects,
stands out clearly from the other distributions. 

It is tempting to directly analyze the distributions $\ell_n$ and to
extract the exponent $\tau$ from the asymptotic behavior of $\ell_n$,
which is algebraic at criticality. However, this method gives far less
accurate results than applying finite-size scaling to observables
involving the sum $\sum_n$ over the cluster sizes $n$.  The main
drawback of the method is the great sensitivity to the location of the
fitting window, i.e., the interval of $n$.  The fitting range cannot be
started at too small cluster sizes, where the distribution has not taken
on its asymptotic form yet, while too large cluster sizes, which are
generated only a few times during a complete Monte Carlo run, are also
to be excluded because of the noise in the data and finite size effects. 
Even in those cases for which we obtained very accurate results through
finite-size scaling analyses, the direct analysis of the distributions
gave unsatisfactory results. 

%\widetext

%

\section{Conclusions}
\label{sec:conclude}
In this paper, the fractal dimensions of spin clusters and their
boundaries appearing in the critical 2D Ising model were studied
numerically.  The Monte Carlo simulations were carried out on
comparatively small lattices.  Standard finite-size scaling was applied
to obtain very precise estimates for the cluster dimensions,
significantly improving existing ones.  The results confirm the exact
theoretical predictions to a high degree of precision. 

For the boundary dimensions, although improving existing estimates, we
obtained less accurate results because of corrections to scaling.  We
observed the strongest corrections for the geometrical hulls, whose
distribution approaches its asymptotic form very slowly.  In a previous
numerical investigation \cite{geoPotts}, where we simulated the hulls of
the spins on the dual lattice directly, corrections to scaling were
virtually absent, allowing us to establish the geometrical hull
dimension to fairly high precision.

To our surprise, we found the fractal dimension of the FK hulls to
converge to the predicted value only when restricting the fitting window
to increasingly smaller lattice sizes.  In general, one expects of
course such a convergence when increasing the lattice size, rather than
decreasing it, so as to minimize corrections to scaling.  The measured
average FK hull size on larger lattices falls below the line
extrapolated from smaller lattices.  The cause for this behavior eludes
us. 

To verify relation (\ref{DH-DEP}), involving the two different boundary
types that can be defined for a cluster, viz.\ hulls and external
perimeters, we treated the two boundary types in a similar manner. 
Usually, hulls are traced out by a directed random walker on the cluster
whereas external perimeters are traced out by a directed random walker
probing the cluster from the outside.  We, on the other hand, applied the
hull algorithm also to the external perimeters of FK clusters, with the
proviso that the random walker can move to a nearest neighbor site on
the FK boundary even when the connecting bond is not set (for the hull,
the bond must be set).

\acknowledgments  
This work is partially supported by the DFG Grant No. JA 483/17-3 and by
the German-Israel Foundation (GIF) under Grant No.\ I-653-181.14/1999. 
A.S. gratefully acknowledges support by the DFG through the
Graduiertenkolleg ``Quantenfeldtheorie'' and the Theoretical Sciences
Center (NTZ) of the Universit\"at Leipzig.  The project is carried out
on a Beowulf GNU/Linux computer cluster.

\end{document}